\documentclass[11pt]{article}
\title{New Trends on N=4 SuperYang-Mills Theory}
\usepackage{mathrsfs}
\usepackage{amsmath}

\global\arraycolsep=1pt
\usepackage[colorlinks=true,backref=true,linkcolor=black,anchorcolor=black,citecolor=black,filecolor=black,menucolor=black,pagecolor=black,urlcolor=black]{hyperref}
\usepackage[Symbol]{upgreek}
\usepackage{bbm}
\usepackage{dsfont}
\usepackage{amssymb}
\usepackage{textcomp}
\newcommand{\ba}{/ \hspace{-2.5mm}}

\newcommand{\scal}[1]{\bigl ({#1} \bigr )}
\def\bea{\begin{eqnarray}}
\def\eea{\end{eqnarray}}
\def\be{\begin{equation}}
\def\ee{\end{equation}}
\newcommand{\CR}{\nonumber \\*}

\newcommand{\trace}{\hbox {Tr}~}

\def\L{{\cal L}}

\def\Lc{\mathscr{L}}

\def\bA{\overset{\circ}{A}}

\usepackage[colorlinks=true,backref=true,linkcolor=black,anchorcolor=black,citecolor=black,filecolor=black,menucolor=black,pagecolor=black,urlcolor=black]{hyperref}

\usepackage[Symbol]{upgreek}
\usepackage{caption}
\usepackage{bbm}
\usepackage{dsfont}
\usepackage{amssymb}
\usepackage{textcomp}
\def\bea{\begin{eqnarray}}
\def\eea{\end{eqnarray}}
\def\be{\begin{equation}}
\def\ee{\end{equation}}

\def\L{{\cal L}}

\def\Lc{\mathscr{L}}

\def\bA{{\overset{\circ}{A}}}
\def\bF{{\overset{\circ}{F}}}

\begin{document}
\allowdisplaybreaks[1]
\renewcommand{\thefootnote}{\fnsymbol{footnote}}

\begin{titlepage}
\begin{flushright}
 LPTHE-05\\
\end{flushright}
\begin{center}
{{\Large \bf 
 New Results on $\mathcal{N}=4$ SuperYang--Mills Theory
 }}
\lineskip .75em
\vskip 3em
\normalsize
{\large L. Baulieu\footnote{email address: baulieu@lpthe.jussieu.fr},
 G. Bossard\footnote{email address: bossard@lpthe.jussieu.fr},
 }\\
$^{*\dagger}$ {\it LPTHE, CNRS and Universit\'es Paris VI - Paris VII, Paris,
France}\footnote{
4 place Jussieu, F-75252 Paris Cedex 05, France.} 
\\
$^{* }$\it Department of Physics and Astronomy, Rutgers University \footnote{
Piscataway, NJ08855-0849,~USA.} 
\\

\vskip 1 em
\end{center}
\vskip 1 em
\begin{abstract}

\end{abstract}
The $\mathcal{N}=4$ SuperYang--Mills theory is covariantly determined by a
$ U(1)\times SU(2) \subset SL(2,\mathds{R})\times SU(2)$ internal symmetry and two scalar 
 and one vector BRST topological symmetry operators.
This determines an off-shell closed sector of $\mathcal{N}=4$ SuperYang--Mills, with 6 generators, which is big enough to fully determine the theory, in a Lorentz covariant way. This reduced algebra derives from horizontality conditions in four
dimensions.   The horizontality conditions  only depend on the geometry of the Yang--Mills fields.   They  also descend from a 
 genuine horizontality condition in eight dimensions. 
 In fact, the $SL(2,\mathds{R})$ symmetry is induced by a dimensional
 reduction from eight to seven dimensions, which establishes a
 ghost-antighost symmetry, while the $ SU(2)$ symmetry occurs by
 dimensional reduction from seven to four dimensions. When the four dimensional manifold is hyperK\"{a}hler, one can perform a twist operation that defines the $\mathcal{N}=4$ supersymmetry and its 
 $SL(2,\mathds{H})\sim SU(4)$ R-symmetry in flat space. (For defining a TQFT on a more general four manifold, one can use 
the internal $ SU(2)$-symmetry and redefine a Lorentz $SO(4)$ invariance). 
These results extend in a covariant way the light cone property that the $\mathcal{N}=4$ SuperYang--Mills theory is actually determined by only 8 independent generators, instead of the 16 generators that occur in the physical representation of the superPoincar\'e algebra. The topological construction disentangles the off-shell closed sector of the (twisted) maximally supersymmetric theory from the (irrelevant) sector that closes only modulo equations of motion. It allows one to escape the question of auxiliary fields in $\mathcal{N}=4$ SuperYang--Mills theory.

\end{titlepage}
\renewcommand{\thefootnote}{\arabic{footnote}}
\setcounter{footnote}{0}



\renewcommand{\thefootnote}{\arabic{footnote}}
\setcounter{footnote}{0}



\section{Introduction}
Recently, we have constructed the genuine $\mathcal{N}=2$
supersymmetric algebra in four and eight dimensions in their twisted
form, directly from extended horizontality conditions
\cite{BBT}. The new features 
were the geometrical construction of both scalar and vector
topological BRST symmetries. 
A remarkable  property  is that the supersymmetry
Yang--Mills algebra, both with 8 and 16 generators, contain an
off-shell closed sector, with 5 and 9 generators, respectively, which
is big enough to completely determine the theory. The key of the
geometrical construction is the understanding that one must determine
the scalar topological BRST symmetry in a way that is explicitly
consistent with reparametrization symmetry. This yields the vector
topological BRST symmetry in a purely geometrical way. 

Here we will extend the result to the case of the maximally
supersymmetric $\mathcal{N}=4$ algebra in 
four dimensions, with its 16 supersymmetric spinorial generators and
$SL(2,\mathds{H})$ internal symmetry\footnote{The internal symmetry
 group $SU(4)$ of $\mathcal{N}=4$ superYang--Mills, defined on a
 Minkowski space, must be replaced on a Euclidean one by the
 Minkowski equivalent $SL(2,\mathds{H})\sim SO(5,1)$ of $SU(4)$. This is implied
 by the fact that $\mathcal{N}=4$ superYang--Mills is the dimensional
 reduction of the ten-dimensional $\mathcal{N}=1$ superYang--Mills
 theory, which is only defined on Minkowski space.}.
The most determining phenomenon occurs when one computes the
dimensional reduction from eight to seven dimensions, which provides an
$SL(2,\mathds{R})$ symmetry. The 
subsequent dimensional reductions to six and four dimensions are then
straightforward, and add a further $\mathds{Z}_2$ and $SU(2)$
internal symmetry, respectively. This explains the organization of the
paper and the eventual obtaining of four-dimensional horizontality
conditions,  which determine the $\mathcal{N}=4$ algebra in a twisted form.
We will also discuss the possible other twists of the supersymmetric theory.

As for the physical application of our construction,  having  obtained  
an off-shell closed algebra
allows one to escape the question of auxiliary fields in
$\mathcal{N}=4$ SuperYang--Mills theory. In a separate publication,
using this algebra and its consequences, we will give an improved
demonstration of the renormalization and finiteness of the  
$\mathcal{N}=4$ superYang--Mills theory  \cite{renor}.



\section{From the $D=8$ to the $D=7 $ topological Yang--Mills theory }

\subsection{Determination of both topological scalar symmetries in seven dimensions}
The $D=8$ topological Yang--Mills theory relies on the following
horizontality equation, completed with its Bianchi identity \cite{BBT}:
 \be 
\label{nero}
( d + s + \delta - i_{\kappa} ) \scal{ A + c + |\kappa| \bar c } + \scal{
 A + c + |\kappa| \bar c }^2 
 = F + \Psi + g(\kappa) \eta + i_{\kappa }
\chi + \Phi + |\kappa|^2 \bar\Phi 
\ee
One has the closure relations:
 \begin{gather}
 s^2 = \delta^2 =0\hspace{10mm} \{s, \delta\} = \L_\kappa
\end{gather}

%
We refer to \cite{BBT} for a detailed explanation of these formula and
the twisted fields that they involve. $\Psi$ is a $1$-form topological
ghost and $\chi$ is an antiselfdual $2$-form in eight dimensions, with 7
independent components. Selfduality exists in eight dimensions when
the manifold has a holonomy group included in $Spin(7)$. The
octonionic invariant $4$-form of such a manifold allows one to define
selfduality, by the decomposition of a 2-form as ${\bf 28}={\bf 7}
\oplus {\bf 21}$. $\kappa$ is a covariantly constant vector, which
exists if the holonomy group is included in $G_2\subset
Spin(7)$. $\Phi$ and $\bar \Phi$ are respectively a topological scalar
ghost of ghost and an antighost for antighost. $c$ and $\bar c$ can be
interpreted as the Faddeev--Popov ghost and antighost of the
eight-dimensional Yang--Mills field $A$. $s$ and $\delta$ are the
scalar and vector topological BRST operators. In flat space, one  can
write $\delta+|\kappa| \delta_{\rm gauge}( \bar c)=\kappa^\mu Q_\mu$, and   we understand that $Q= s +
\delta_{\rm gauge}(c)$ and
$Q_\mu$ count for 9 independent generators, giving an off-shell
closed sector of $\mathcal{N}=2,D=8$ twisted supersymmetry, which fully
determines the theory \cite{BBT}. 

To determine  the topological symmetry in seven dimensions, we
start  from  a  manifold in eight dimensions  that is reducible, $M_8
=N_7\times S^1$, where $N$ is a $G_2$-manifold. We can chose
$\kappa$ as a tangent vector to $S^1$. Thus, the 8-dimensional vector symmetry along the
circle reduces to a scalar one in seven dimensions, $\bar s$, and the
reduction of the antiselfdual $2$-form $\chi$ gives a
seven-dimensional $1$-form $\bar \Psi$. So, the dimensional reduction
of the horizontality condition (\ref{nero}) is
\be 
\label{ddr}
(d + s + \bar s) \scal{ A + c + \bar c} + \scal{A + c + \bar c}^2 = F
+ \Psi + \bar \Psi + \Phi + L + \bar \Phi 
\ee
Indeed, with our choice of $\kappa$,  $L =i_\kappa A
=A_8 $ in eight dimensions. Eq.~(\ref{ddr}) can be given    a different
interpretation than Eq.~(\ref{nero}). It contains    no   vector symmetry
and looks like a standard 7-dimensional  BRST-antiBRST
equation. The transformation of $L$, the origin of which is
$A_8$, is now given by the Bianchi identity of Eq.~(\ref{ddr}).  In
fact, after dimensional reduction, $L$ is understood as a curvature. 
Using the convenient pyramidal diagrammatic description of
ghost-antighost structures, we can rewrite    the
 field description, as follows: 
\be \begin{array}{ccccc}
\hspace{8mm} & \hspace{5mm} & A\,,\,L & \hspace{5mm} & \hspace{8mm}
\\&&&&\\
 & \Psi\,,\,\bar \eta & & \,\,\,\,\bar\Psi \,\,\,\,& \\&&&&\\
\Phi & & & & \bar \Phi \\&&&&\\
 & & & \eta & 
\end{array} \hspace{7mm} \longrightarrow \hspace{7mm} \begin{array}{ccccc}
\hspace{5mm} & \hspace{10mm} &\,\, A\,\, & \hspace{10mm} & \hspace{5mm}
\\&&&&\\
 & \Psi & & \bar\Psi & \\&&&&\\
\Phi & & L & & \bar \Phi \\&&&&\\
 &\bar \eta & & \eta & 
\end{array} \ee 

As compared to the   asymmetrical diagram on the left hand-side, the  
 one on the right hand-side exhibits an $SL(2,\mathds{R})$ symmetry, which
counts the ghost antighost numbers. 
Indeed, each line of this diagram corresponds to an irreducible
representation of $SL(2,\mathds{R})$, namely a completely symmetrical
$SL(2,\mathds{R})$-spinorial tensor with components $\phi ^{g, G-
 g}$, where $g$ and $G-g$ are respectively the (positive) ghost
  and antighost numbers of $\phi$, and $2 g- G$ is the effective
ghost number of $\phi$. This $SL(2,\mathds{R})$ symmetry actually
applies to the covariant ghost-antighost spectrum of a $p$-form gauge
field $\phi_p$, $\tilde \phi_p= \sum_{ 0\leq G \leq p} \sum_{0\leq g \leq G}
\phi_{p-G}^{g, G-g}$.

In fact, the fields $(\Psi, \bar \Psi)$ and $(\eta,\bar \eta) $ can be
identified as $SL(2,\mathds{R})$ doublets, $\Psi^\alpha$ and $\eta^\alpha$, 
$\alpha=1,2$,
and the  three scalar fields  $(\Phi, L, \bar \Phi)$ as a $SL(2,\mathds{R})$ triplet, $
\Phi^i$, $ i=1,2,3$. The index $\alpha$ and $i$ are
respectively raised and lowered   by the volume form 
$\varepsilon_{\alpha\beta}$ 
of
$SL(2,\mathds{R})$ and the Minkowski metric $\eta_{ij}$ of signature
$(2,1)$. Both BRST and antiBRST operators  can be assembled into a $SL(2,\mathds{R})$ doublet $s^\alpha=(s,\bar s)$. 

The horizontality condition (\ref{ddr}) can be solved, with the
introduction of three 0-form Lagrange multipliers, $\eta, \bar \eta, b$ and a 1-form
$T$. 
\be\label{s} \begin{array}{rclcrcl}
sA &=& \Psi - d_A c & \hspace{10mm}& \bar s A &=& \bar \Psi - d_A
\bar c \\*
s \Psi &=& - d_A \Phi - [c ,\Psi] & & \bar s \Psi &=& -T - d_A L -
[\bar c , \Psi ]\\*
s \Phi &=& -[c , \Phi] & & \bar s \Phi &=& - \bar \eta- [\bar c ,
\Phi] \\
s \bar \Phi &=& \eta- [c, \bar \Phi] & & \bar s \Phi &=& - [\bar c ,
\bar \Phi]\\*
s \eta &=& [\Phi, \bar \Phi] - [c , \eta] & & \bar s \eta &=& - [\bar
\Phi , L] - [\bar c , \eta ] \\*
s L &=& \bar \eta- [c, L] & & \bar s L &=& - \eta - [\bar c , L]\\*
s \bar \eta &=& [\Phi, L]- [c, \bar \eta] & & \bar s \bar \eta &=& [\Phi, \bar
\Phi]- [\bar c , \bar \eta]\\*
s \bar \Psi &=& T- [c , \bar \Psi] & & \bar s \bar \Psi &=& - d_A \bar
\Phi - [\bar c , \bar \Psi] \\*
s T &=& [\Phi, \bar \Psi ]- [ c, T]& & \bar s T &=& - d_A \eta + [L,
\bar \Psi] - [\bar \Phi, \Psi] - [\bar c , T] \\*\\
s c &=& \Phi - c^2 & & \bar s c &=& L - b \\*
s \bar c &=& b- [c, \bar c] & & \bar s \bar c &=& \bar \Phi - {\bar
 c}^2\\*
s b &=& [\Phi, \bar c] - [c, \bar c] & & \bar s b &=& \eta + [\bar c ,
L]
\end{array}\ee
 The transformation laws of the  $b$  field cannot be made however $SL(2,\mathds{R})$
covariant. Its introduction allows one to raise the degeneracy arising
from the equation $s \bar c +\bar s c +[c,\bar c]=L$, which basically selects a ghost direction, and freezes the $SL(2,\mathds{R})$ symmetry.
In fact, going from a Cartan to a nilpotent algebra breaks this
$SL(2,\mathds{R})$, and in this way, defines the net ghost number symmetry from the
residual unbroken $U(1) \subset SL(2,\mathds{R})$. The Cartan
algebra (that we will denote by the subindex $\scriptstyle (c)$)
is on the other hand fully $SL(2,\mathds{R})$ covariant. It is
obtained by adding gauge transformations with parameters $c$ and
$\bar c$, from $s$ and $\bar s$, respectively. It reads as: 
\be
\begin{split}
s_{\scriptscriptstyle (c)}^\alpha A &= \Psi^\alpha \\*
s_{\scriptscriptstyle (c)}^\alpha \Psi_\beta &= \delta^\alpha_\beta T -
{{\sigma^i}_\beta}^\alpha d_A \Phi_i \\*
s_{\scriptscriptstyle (c)}^\alpha \Phi_i &= \frac{1}{2} {\sigma_i}^{\alpha\beta} \eta_\beta
\end{split}
\hspace{10mm}
\begin{split}
s_{\scriptscriptstyle (c)}^\alpha \eta_\beta &= - 2
{{\sigma^{ij}}_\beta}^\alpha [ \Phi_i, 
\Phi_j] \\*
s_{\scriptscriptstyle (c)}^\alpha T &= \frac{1}{2} d_A \eta^\alpha + \sigma^{i\,
 \alpha\beta}[\Phi_i, \Psi_\beta]
\end{split}
\ee
 The equivariant (Cartan) algebra is the one that will match by twist with the relevant part of the twisted supersymmetry algebra.
 Its closure is only modulo gauge transformations, with parameters that are ghosts of ghosts.
 \subsection{Determination of the equivariant part of both topological
 vector symmetries in seven dimensions} 
 
We have produced by dimensional reduction a new scalar (antiBRST)
topological symmetry operator. However, we have apparently lost the
rest of the vector symmetry in eight dimensions, since we have chosen
$\kappa$ along the circle of dimensional reduction. The
freedom of choosing this circle 
generates an automorphism of the seven dimensional
symmetry. It allows one to obtain two vector topological
symmetries, in seven dimensions, in a $SL(2,\mathds{R})$ symmetric
way. 

In the genuine theory in eight dimensions, the definition of a
chosen $Spin(7)$ structure is actually arbitrary. The different choices of a
$Spin(7)$ structure on the eight dimensional Riemann manifold can be
parametrized by the   transformations of  $SO(8)\setminus
Spin(7)$. However, such   transformations cannot
 be represented on   eight dimensional fields such as the  antiselfdual 2-form $\chi_{\mu\nu}$. 
They are 
  generated by antiselfdual 2-form infinitesimal parameters, which 
 can be parametrized by seven-dimensional
vectors, after dimensional reduction to 7 dimensions.  One can  thus
define  the following  (commuting)
7-dimensional  derivation $\upgamma$, which depends on a covariantly constant
seven-dimensional vector $\kappa^\mu$, (the indices $\mu, \nu...$ are seven dimensional), and acts as follows on the  7-dimensional fermion fields:
\bea
\upgamma \Psi^\alpha_\mu &=& - \kappa_\mu \eta^\alpha -
{C_{\mu\nu}}^\sigma \kappa^\nu \Psi^\alpha_\sigma \CR
\upgamma \eta^\alpha &=& \kappa^\mu \Psi^\alpha_\mu
\eea
The action of $\upgamma $ is  zero on the bosonic field of the equivariant BRST algebra, but  on the Lagrange multiplier field $T$. (We will shortly define    $\upgamma T$, by consistency).
 $C_{\mu\nu\sigma}$ is the seven dimensional $G_2$ invariant octonionic $3$-form and its dual is the $4$-form $C ^\star _{\mu\nu\sigma\rho} $. We will use the notation
 $i_\kappa \star C ^\star w_1 = -{C_{\mu\nu}}^\sigma \kappa^\nu w_\sigma dx^\mu$.
The  derivation   $\upgamma$    expresses the arbitrariness in the choice of an eighth
component,  in order to perform the dimensional reduction. 
To each
constant vector on $N$, one  can assign a  $U(1)$ group, which is subset of
$SO(8) \setminus Spin(7)$. Since $\{ s_{\scriptscriptstyle (c)}^\alpha , s_{\scriptscriptstyle (c)}^\beta \}
=\sigma^{i\,\alpha \beta}
\delta _{\rm gauge} (\Phi_i)$, one can verify  on all fields:
\be \{e^{t \upgamma} s_{\scriptscriptstyle (c)}^\alpha e^{- t \upgamma} , e^{t \upgamma}
s_{\scriptscriptstyle (c)}^\beta e^{- t \upgamma}\} = \{ s_{\scriptscriptstyle (c)}^\alpha , s_{\scriptscriptstyle (c)}^\beta \}\ee
One defines the vector operator:
\be \delta_{\scriptscriptstyle (c)}^{\alpha} \equiv [s_{\scriptscriptstyle (c)}^\alpha, \upgamma]\ee 
Since $[[s_{\scriptscriptstyle
 (c)}^\alpha , \upgamma], \upgamma] = -s_{\scriptscriptstyle
 (c)}^\alpha$, one has: \be\label{auto}
 e^{t \upgamma} s_{\scriptscriptstyle (c)}^\alpha e^{- t \upgamma} =
 \cos t \,\,s_{\scriptscriptstyle (c)}^\alpha - \sin
 t\,\,\delta_{\scriptscriptstyle (c)}^\alpha\ee 
To ensure the validity of this   formula on $T$, one    defines the transformation of the auxiliary field $T$ as follows:
\be \upgamma T =- i_\kappa F - 2 i_\kappa \star C^\star T \ee

Computing the
commutators of $s^\alpha _{\scriptscriptstyle (c)} $ and $\upgamma$, one gets 
  the action of $ \delta_{\scriptscriptstyle (c)}^\alpha$:
\bea
\delta_{\scriptscriptstyle (c)}^\alpha A&=& g(\kappa) \eta^\alpha + i_\kappa \star C^\star
\Psi^\alpha \CR
\delta_{\scriptscriptstyle (c)}^\alpha \Psi_\beta &=& \delta^\alpha_\beta i_\kappa \scal{ F +
 \star C^\star T} + {{\sigma^i}_\beta}^\alpha i_\kappa \star C^\star
d_A \Phi_i - 2 g(\kappa) {{\sigma^{ij}}_\beta}^\alpha [\Phi_i, \Phi_j]\CR
\delta_{\scriptscriptstyle (c)}^\alpha \Phi_i &=& - \frac{1}{2} {\sigma_i}^{\alpha\beta}
i_\kappa \Psi_\beta \\*
\delta_{\scriptscriptstyle (c)}^\alpha \eta_\beta &=& - \delta^\alpha_\beta i_\kappa T +
{{\sigma^i}_\beta}^\alpha \Lc_\kappa \Phi_i \CR
\delta_{\scriptscriptstyle (c)}^\alpha T&=& \frac{1}{2} d_A i_\kappa \Psi^\alpha - i_\kappa
\star C^\star d_A \eta^\alpha + g(\kappa) \sigma^{i\, \alpha\beta}
[\Phi_i, \eta_\beta] -\sigma^{i\, \alpha\beta} i_\kappa \star C^\star
[\Phi_i, \Psi_\beta ]- \Lc_\kappa \Psi^\alpha \nonumber
\eea
 $\delta^\alpha_{\scriptscriptstyle (c)}$ is a pair of two vector
 symmetries in seven dimensions, which transform as an
 $SL(2,\mathds{R})$-doublet. This completes the scalar doublet
 $s_{\scriptscriptstyle (c)}^\alpha$. 
 
 Then, one can verify that the anticommutation relations for the vector operator $ \delta_{\scriptscriptstyle (c)}^\alpha$ are:
\be \label{clo} \{ s_{\scriptscriptstyle (c)}^\alpha , \delta_{\scriptscriptstyle
 (c)}^\beta \} = \varepsilon^{\alpha\beta} 
\scal{ \L_\kappa + \delta_{\mathrm{gauge}}(i_\kappa A)} \hspace{10mm}
 \{ \delta_{\scriptscriptstyle (c)}^\alpha
 ,\delta_{\scriptscriptstyle (c)}^\beta \} = 2
 \sigma^{i\,\alpha\beta} \delta_{\mathrm{gauge}}(\Phi_i) \ee
Reciprocally, these closure relations uniquely determine $\delta^\alpha
_{\scriptscriptstyle (c)}$, from the knowledge of $s^\alpha
_{\scriptscriptstyle (c)}$.
 
In the next section, we will re-derive these transformation laws, from horizontality equations. Moreover, we will  extend them as  nilpotent transformations, by including gauge transformations. 

It is   instructive to check the expression we have just obtained for  $ \delta_{\scriptscriptstyle (c)}^\alpha$, by starting from Eq.~(\ref{nero}) in 8 dimensions (with the notational change  $\bar c  \to \gamma$), and computing the dimensional reduction with a vector $\kappa$ along $N$, instead of along the
circle. This gives:
\bea\label{delta} \begin{split}
\delta A &= g(\kappa) \eta + i_\kappa \star C^\star \bar \Psi -
|\kappa| d_A \gamma \\*
\delta \Psi &= i_\kappa \scal{ F - \star C^\star T} + g(\kappa) [\Phi,
\bar \Phi] - [|\kappa| \gamma, \Psi] \\*
\delta \Phi &= i_\kappa \Psi - [|\kappa| \gamma, \Phi] 
\end{split} \hspace{10mm} \begin{split}
\delta \bar \Phi &= - [|\kappa| \gamma, \bar \Phi] \\*
\delta \eta &= \Lc_\kappa \bar \Phi - [|\kappa|\gamma, \eta]\\*
\delta L &= i_\kappa \bar \Psi - [|\kappa|\gamma, L]\\*
\delta \bar \eta &= i_\kappa \scal{ T + d_A L} - [|\kappa| \gamma,
\bar \eta] \end{split} \CR
\delta \bar \Psi = i_\kappa \star C^\star d_A \bar \Phi - g(\kappa)
[\bar \Phi, L] - [|\kappa| \gamma, \bar \Psi] \CR
\delta T = \Lc_\kappa \bar \Psi + i_\kappa \star C^\star \scal{ d_A
 \eta + [\bar \Phi]} + g(\kappa) [ L, \eta] - g(\kappa) [ \bar \Phi,
\bar \eta] - [|\kappa| \gamma, T]
\eea
One can then verify: 
\be [ \bar s , \upgamma ] = \delta \ee
with the modified  definition that 
\begin{gather}
\upgamma T = i_\kappa F - 2 i_\kappa \star C^\star T - i_\kappa \star
C^\star d_A L \CR
\upgamma \bar c = - |\kappa| \gamma \hspace{10mm} \upgamma \gamma =
\frac{1}{|\kappa|} \bar c 
\end{gather}
The difference between this expression of $\delta+ |\kappa|
\delta_{\rm gauge}(\gamma)$ and that of a component of the 
$SL(2,\mathds{R})$ covariant vector symmetry operators
$\delta^\alpha_{\scriptscriptstyle (c)}$ is just a field
redefinition.


\subsection{The complete Faddeev--Popov ghost dependent vector and
 scalar topological symmetries in seven dimensions} 
We now directly construct the   scalar and vector BRST topological  operators
$s^\alpha$ and $\delta^\alpha$, the equivariant 
analogs of which are $s_{\scriptscriptstyle (c)}^\alpha$ and $\delta_{\scriptscriptstyle (c)}^\alpha$.
One needs  scalar Faddeev--Popov ghosts,
$c,\bar c,\gamma,\bar \gamma$, which are associated to the 
equivariant BRST operators $s_{\scriptscriptstyle (c)},\bar
s_{\scriptscriptstyle (c)},\delta_{\scriptscriptstyle (c)},\bar
\delta_{\scriptscriptstyle (c)}$, respectively. 

The   relations (\ref{clo}) suggests the following horizontality condition, with $( d + s + \bar s + \delta + \bar\delta)^2=0$:
\begin{multline}\label{magics}
( d + s + \bar s + \delta + \bar\delta) \scal{ A+ c + \bar c +
 |\kappa| \gamma + |\kappa| \bar \gamma } + \scal{ A+ c + \bar c +
 |\kappa| \gamma + |\kappa| \bar \gamma }^2 \\*= F + \Psi + \bar\Psi
+ g(\kappa) \scal{ \eta + \bar \eta} +
i_\kappa \star C^\star \scal{\Psi + \bar\Psi} + ( 1+ |\kappa|^2)
\scal{ \Phi + L + \bar\Phi } 
\end{multline}
It is $SL(2, \mathds{R})$ and $\upgamma$ invariant. 
By construction, this equation has the following indetermination: 
\be 
\{ s, \delta \} + \{ \bar s , \bar \delta\}= 0
\ee
This degeneracy is raised, owing to the introduction of the constant vector $\kappa$, with,
\be
\{ s,\delta\} =\L_\kappa
\quad\quad 
 \{\bar s, \bar \delta \} =-\L_\kappa
\ee
This relation is fulfilled by completing Eq.~(\ref{magics}) by the
following ones: 
\bea \label{magic+} 
( d + s + \delta - i_\kappa) \scal{ A+ c +
 |\kappa| \gamma } + \scal{ A+ c + 
 |\kappa| \gamma } ^2 \hspace{50mm}\CR \hspace{25mm}= F + \Psi 
+ g(\kappa) \eta +
i_\kappa \star C^\star \scal{ \bar\Psi} + 
\scal{ \Phi + |\kappa|^2 \bar\Phi } 
\\*
\label{magic-} 
( d + \bar s + \bar \delta +i_\kappa) \scal{ A+\bar c +
 |\kappa| \bar \gamma } + \scal{ A+ \bar c + 
 |\kappa| \bar \gamma } ^2 \hspace{50mm}\CR \hspace{25mm} = F +
\bar \Psi + g(\kappa) \bar \eta +
i_\kappa \star C^\star \scal{ \Psi} + 
\scal{ \bar \Phi + |\kappa|^2 \Phi } 
\eea
The $SL(2,\mathds{R})$ and $\upgamma$ invariant Eqs.~(\ref{magics}),(\ref{magic+}),(\ref{magic-}) are consistent, but not independent. Only one of 
Eqs.~(\ref{magic+}) or (\ref{magic-}) is needed to complete
Eq.~(\ref{magics}). 
To solve equations of the type $s
 \bar c+\bar s c + ...=0$, one reduces the $SL(2,\mathds{R})$ to a $U(1)$ symmetry, with $s
 \bar c =b+...$. One can verify that, by expansion in ghost number, the horizontality conditions reproduce the transformations (\ref{s}) and (\ref{delta}). 
 
 We note that the number of symmetries carried by the generators
 $s_{\scriptscriptstyle (c)}^\alpha$ and $\delta_{\scriptscriptstyle (c)}^\alpha$ is 1+1+7+7=16. They yield, by untwisting in flat space, 
the complete set of Poincar\'e supersymmetry generators (we donnot reproduce here this lengthy computation).
We can
 understand the seven generators determined by $\bar \delta$ as the
 dimensional reduction of the twisted antiselfdual generator of
 $\mathcal{N}=2, D=8$ supersymmetry, and $\bar s$ and $ \delta$ as the
 dimensional reduction of the twisted vector 
 supersymmetry in 8 dimensions. Only a maximum of 9=1+1+7 generators,
 among the 16 ones  that are determined by $s,\bar s, \delta, \bar \delta$,
 build an off-shell closed 
 algebra, since both operators  $\delta_{\scriptscriptstyle
   (c)}^\alpha$ depend on a single vector 
 $\kappa$. In fact, the commutation relations of the vector
 operators $Q_\mu$ and $\bar Q_\nu$, where $\delta_{\scriptscriptstyle
   (c)}=\kappa^\mu Q_\mu $ 
 and  $\bar \delta_{\scriptscriptstyle (c)}= \kappa^\mu \bar Q_\mu$,
  yield non closure terms for their
 antisymmetric part in $\mu,\nu$. One can choose $Q,\bar Q, Q_\mu$ as
 such a maximal  subalgebra.

Dimensional reduction   therefore transforms the $\mathcal{N}_T = 1$
eight-dimensional theory into a $\mathcal{N}_T = 2$ theory, with an
$SL(2,\mathds{R})$ internal symmetry, and a $G_2\subset Spin(7)$ Lorentz symmetry. As a matter of fact, this algebra gives the
$SL(2,\mathds{R})$ invariant twisted supersymmetry transformations
\cite{glouk} in the limit of flat manifold. 

\subsection{Seven-dimensional invariant action}

The most general gauge invariant topological gauge function $\Uppsi $, 
 which yields a $\delta$ invariant  action $S=s \Uppsi - \frac{1}{2} \int_M C_{\, \wedge}
 \trace F_{\, \wedge} F$, is:
\be
 \Uppsi = 2 \int_M \trace \biggl( C^\star_{\, \wedge}
\bar \Psi_{\, \wedge} F + \bar \Psi \star \scal{ d_A L + T} + \Psi
\star d_A \bar \Phi + \star \eta [\Phi, \bar \Phi] + \star \bar \eta
[\bar \Phi, L] \biggr) 
\ee 
 This function   turns out to be $\bar s$-exact and thus $\bar s$
 invariant. Moreover, $S$ is  $\bar \delta $ invariant.  By
 using the $SL(2,\mathds{R})$ covariant form of the algebra, we can
 compute the gauge function $ \Uppsi $  in a manifestly invariant way, as a
 component of a doublet $ \Uppsi_\alpha$. One has $\sigma^{i\,
   \alpha\beta} \delta_{{\scriptscriptstyle (c)}\, \alpha}
 \Uppsi_\beta = 0 $ and $\Uppsi _\alpha= \delta _{(c) \alpha}
 \mathscr{G}$. $S=s_{\scriptscriptstyle (c)}^\alpha \Uppsi_\alpha- \frac{1}{2} \int_M C_{\, \wedge} \trace F_{\, \wedge} F$,
 and, with our conventions, $\Uppsi \propto \Uppsi _1$.
The automorphism generated by $\upgamma$ leaves invariant neither the
gauge function, nor the action, since its action breaks the
$Spin(7)$-structure of the 8-dimensional theory. These properties will remain analogous in 4 dimensions, and we will give more details in this case.

\section{Reduction to four dimensions and $\mathcal{N}=4$ theory}

The process of dimensional reduction can be further done, from 7 dimensions. The ghost antighost symmetry that has   appeared when going down from 8 to 7 dimensions will continue to hold true, and therefore, one 
remains in the framework of an $\mathcal{N}_T = 2$ theory, with $SL(2,\mathds{R})$ invariance.

 One  is 
concerned by going down from seven to four dimensions. The
$SO(7)$-symmetry is decomposed into $SO(4)\times SO(3)\sim Spin(4)\times
SU(2)$, and insides this decomposition, the $G_2$-symmetry is
decomposed into $SU(2)\times SU(2)$. So the (twisted) 4-dimensional
theory is expected to have $SU(2)\subset Spin(4)$ Lorentz symmetry,
with an $SL(2,\mathds{R})\times SU(2)$ internal
symmetry. Thus, starting from a manifold of holonomy included in $G_2$
suggests that the theory will be defined on a  4-dimensional
hyperK\"{a}hler manifold. 

In fact, by identifying the $SU(2)$ part of the internal symmetry
group as the missing chiral $SU(2)$ part of the $Spin(4)$-invariance,
one can eventually restore the full $Spin(4)$ invariance,
 and define the
topological theory on more general manifolds. It is useful to
use  a hyperK\"{a}hler structure   to simplify the form
of the equations. For instance, given the 3 constant
hyperK\"{a}hler 2-forms $J^I_{\mu\nu}$, the antiselfdual $2$-form
$\chi_{\mu\nu}$ can be written as $\chi_{\mu\nu} = \chi_I
J^I_{\mu\nu}$, where $\chi_I$ is a $SU(2)$ triplet made of
scalars. (Capital indices as
$I$ are devoted to the adjoint representation of the $SU(2)$
symmetry, the scalars $h^I$ correspond to $A_7,A_6, A_5$, etc...). This allows simplified expressions for scalars, such as, for instance, 
$\varepsilon_{IJK} h^I h^J h^K$ instead of   $ {h_\mu}^\nu {h_\nu}^\sigma {h_\sigma}^\mu$. 
Moreover, one is  interested  in obtaining by twist the $\mathcal{N}=4$
superYang--Mills theory. This is a further justification of the restricted choice   of a 
hyperK\"{a}hler manifold, since two constant spinors are needed to
perform the twist operation and eventually to map the topological
ghosts on spinors.


\subsection{Equivariant scalar and vector algebra in four dimensions }
By  dimensional reduction of  the  seven-dimensional equations of
section 2, one  can compute the Cartan 
$SL(2,\mathds{R})$ doublet of scalar topological BRST operators for
the topological symmetry in four dimensions: 
\bea
\label{BRST} \begin{split}
s_{\scriptscriptstyle (c)}^\alpha A &= \Psi^\alpha \\*
s_{\scriptscriptstyle (c)}^\alpha \Psi_\beta &= \delta^\alpha_\beta T -
{{\sigma^i}_\beta}^\alpha d_A \Phi_i \\*
s_{\scriptscriptstyle (c)}^\alpha \Phi_i &= \frac{1}{2} {\sigma_i}^{\alpha\beta} \eta_\beta \\*
s_{\scriptscriptstyle (c)}^\alpha \eta_\beta &= - 2 {{\sigma^{ij}}_\beta}^\alpha [\Phi_i,
\Phi_j] \\*
s_{\scriptscriptstyle (c)}^\alpha T &= \frac{1}{2} d_A \eta^\alpha + \sigma^{i\, \alpha \beta}[\Phi_i,
\Psi_\beta] \end{split} \hspace{10mm} \begin{split}
s_{\scriptscriptstyle (c)}^\alpha h^I &= \chi^{\alpha\, I} \\*
s_{\scriptscriptstyle (c)}^\alpha \chi^I_\beta &= \delta^\alpha_\beta H^I +
{{\sigma^i}_\beta}^\alpha [\Phi_i, h^I] \\*
s_{\scriptscriptstyle (c)}^\alpha H^I &= \frac{1}{2} [\eta^\alpha, h^I] + \sigma^{i\, \alpha\beta}
[\Phi_i, \chi^I_\beta]\end{split}
\eea
One has  the closure relation $s_{\scriptscriptstyle (c)}^{\{\alpha}s _{\scriptscriptstyle (c)}^{\beta\}} = \sigma^{i\,
 \alpha\beta} \delta_{\mathrm{gauge}}(\Phi_i)$. The Cartan vector algebra is: 
\bea
\delta_{\scriptscriptstyle (c)}^\alpha A &=& g(\kappa) \eta^\alpha + g(J_I \kappa)
\chi^{\alpha\, I} \CR
\delta_{\scriptscriptstyle (c)}^\alpha \Psi_\beta &=& \delta^\alpha_\beta \scal{i_\kappa F -
 g(J_I \kappa) H^I} + {{\sigma^i}_\beta}^\alpha g(J_I \kappa)
[\Phi_i, h^I] - 2 {{\sigma^{ij}}_\beta}^\alpha g(\kappa) [\Phi_i ,
\Phi_j] \CR
\delta_{\scriptscriptstyle (c)}^\alpha \Phi_i &=& -\frac{1}{2} {\sigma_i}^{\alpha\beta}
i_\kappa \Psi_\beta \CR
\delta_{\scriptscriptstyle (c)}^\alpha \eta_\beta &=& - \delta^\alpha_\beta
i_\kappa T + {{\sigma^i}_\beta}^\alpha \Lc_\kappa \Phi_i \CR
\delta_{\scriptscriptstyle (c)}^\alpha T &=& \frac{1}{2} d_A i_\kappa
\Psi^\alpha - g(J_I \kappa) \scal{ [\eta^\alpha, h^I] + \sigma^{i\, \alpha\beta} [\Phi_i, \chi^I_\beta]} + g(\kappa) \sigma^{i\, \alpha\beta}
[\Phi_i, \eta_\beta] - \Lc_\kappa \Psi^\alpha \CR
\delta_{\scriptscriptstyle (c)}^\alpha h^I &=& - i_{J^I \kappa} \Psi^\alpha \CR
\delta_{\scriptscriptstyle (c)}^\alpha \chi^I_\beta &=& \delta^\alpha_\beta \scal{ \Lc_\kappa
 h^I + i_{J^I \kappa} T} + 
{{\sigma^i}_\beta}^\alpha \Lc_{J^I \kappa} \Phi_i \CR
\delta_{\scriptscriptstyle (c)}^\alpha H^I &=& \frac{1}{2} [ i_\kappa \Psi^\alpha,h^I] +
\Lc_{J^I \kappa} \eta^\alpha + \sigma^{i\, \alpha\beta}
[\Phi_i, i_{J^I \kappa} \Psi_\beta] - \Lc_\kappa \chi^{\alpha\, I}
\eea
One has, $\delta_{\scriptscriptstyle (c)}^{\{\alpha} \delta_{\scriptscriptstyle (c)}^{\beta \}} = |\kappa|^2
\sigma^{i\, \alpha\beta} \delta_{\mathrm{gauge}}(\Phi_i)$,  and  $\delta_{\scriptscriptstyle (c)}^\alpha$
anticommute with $s_{\scriptscriptstyle (c)}^\alpha$,  as follows\footnote{Note that, the existence of a covariantly constant vector field on a
hyperK\"{a}hler manifold $M$  implies  that $M$ is flat. This explains the possible use 
 of a complete quaternionic base of the
tangent space $TM$: $\kappa$, $J^I \kappa$ for the description of the vector symmetry.}: 
\be
\{s_{\scriptscriptstyle (c)}^\alpha , \delta_{\scriptscriptstyle (c)}^\beta
\} = \varepsilon^{\alpha\beta} \scal{
 \L_\kappa + \delta_{\mathrm{gauge}}(i_\kappa A)}
\ee
The four dimensional vector operators are $s^{\scriptscriptstyle
 \alpha}$\hspace{-1mm}-exact, as in seven dimensions, $\delta_{\scriptscriptstyle (c)}^\alpha = [s_{\scriptscriptstyle (c)}^\alpha , \upgamma]
$, where the non zero component of $\upgamma$ are 
\be \begin{split}
\upgamma \Psi_\alpha &= - g(\kappa) \eta_\alpha - g(J_I \kappa)
\chi_\alpha^I \\*
\upgamma \eta_\alpha &= i_\kappa \Psi_\alpha \\*
\upgamma \chi_\alpha^I &= i_{J^I \kappa} \Psi_\alpha
\end{split}\hspace{10mm} \begin{split}
\upgamma T &= - i_\kappa F + 2 g(J_I \kappa) H^I \\*
\upgamma H^I &= - \Lc_\kappa h^I - 2 i_{J^I \kappa} T \end{split}
\ee
As in seven dimensions, one has a $U(1)$ automorphism of the algebra,
which is not a symmetry of the theory, with 
$
e^{ t \upgamma} s_{\scriptscriptstyle (c)}^\alpha e^{- t \upgamma} = \cos t\,\,
s_{\scriptscriptstyle (c)}^\alpha - \sin t \,\, \delta_{\scriptscriptstyle (c)}^\alpha$ and $ e^{t \upgamma}
\delta_{\scriptscriptstyle (c)}^\alpha e^{- t \upgamma} = \cos t\,\,
\delta_{\scriptscriptstyle (c)}^\alpha + \sin t \,\, s_{\scriptscriptstyle (c)}^\alpha
$.
\subsection{Invariant action}
There are two gauge functions
which fit in a fundamental multiplet of $SL(2,
\mathds{R})$, and satisfy: 
\be \label{guil} \sigma^{i\, \alpha\beta} \delta_{{\scriptscriptstyle (c)}\, \alpha} \Uppsi_\beta = 0 \ee
The action is defined as:
\be\label{action}
S = - \frac{1}{2} \int_M \trace F_{\, \wedge} F + s_{\scriptscriptstyle (c)}^\alpha
\Uppsi_\alpha 
\ee
Eq.~({\ref{guil}) completely constrains the gauge function (up to a global
scale factor), as follows: 
\begin{multline}
\Uppsi_\alpha = \int_M \trace \biggl( \star \chi^I_\alpha H_I + 
\chi^I_\alpha J_I \star F - \Psi_\alpha \star T \biggr . 
+ J^I \star
\Psi_{\alpha\, \wedge} d_A h_I - {{\sigma^i}_\alpha}^\beta \Psi_\beta
\star d_A \Phi_i \\*- 2 \star {{\sigma^{ij}}_\alpha}^\beta \eta_\beta
[\Phi_i, \Phi_j] - \star {{\sigma^i}_\alpha}^\beta \chi^I_\beta
[\Phi_i, h_I] +\frac{1}{2} \star \varepsilon_{IJK} \chi^I_\alpha [h^J, h^K]
\biggr)
\end{multline}

The action (\ref{action}) is $\delta ^ \alpha$ and 
$s^ \alpha$ invariant. Indeed, one can check that it verifies: 
\be S = - \frac{1}{2} \int_M \trace F_{\,\wedge}
F + s_{\scriptscriptstyle (c)}^\alpha s_{{\scriptscriptstyle (c)}\,
  \alpha} \mathscr{F} \,\, =
- \frac{1}{2} \int_M \trace F_{\, \wedge} F + s_{\scriptscriptstyle (c)}^\alpha \delta_{{\scriptscriptstyle (c)}\,\alpha}
\mathscr{G}
\ee
with
 \be 
 \mathscr{F} = \int_M \trace \biggl( \star h_I H^I + 
h_I J^I \star F + \frac{1}{3} \varepsilon_{IJK} h^I h^J h^K 
 - \frac{1}{2} \Psi^\alpha \star \Psi_\alpha + \frac{1}{2}\star \eta^\alpha
 \eta_\alpha \biggr) 
\ee
\vspace{-5mm}
 \begin{multline}
\mathscr{G}
= \int_M \trace \biggl( \,\,- \frac{1}{2} g(\kappa)_{\,
 \wedge} \scal{ 
 ( A - \bA)_{\, \wedge} ( F + \bF) - {\textstyle \frac{1}{3}} ( A -
 \bA)^3 } \biggr . \\*- \frac{1}{2} \star \varepsilon_{IJK} h^I
\Lc_{J^J \kappa} h^K + \star \, s_{\scriptscriptstyle (c)}^\alpha \delta_{{\scriptscriptstyle (c)}\,\alpha} \, \, \scal{ \, 
{\textstyle  \frac{1}{2}} \, h_I h^I\, - \, {\textstyle \frac{2}{3}} \, \Phi^i \Phi_i\, }
\biggr) \hspace{1cm} 
\end{multline}

These facts remind us that we are in the context of a $\mathcal{N}_T=2$ theory. The critical points of the Morse function $ \mathscr{F} $ in the field space are
given by the equations
\begin{gather}
 J^I \star F + \frac{1}{2} \star {\varepsilon^I}_{JK} [h^J, h^K] =
 0 \CR
 d_A \star h_I J^I = 0
  \label{MP}
\end{gather}
Eqs. (\ref{MP})  are the dimensional reduction of selfduality
equations in 7 dimensions. 
 \cite{Vafa,MQ,wang} display analogous equations, corresponding to the  
 dimensional reduction of the selfduality equation in 8 dimensions.  \cite{dijk}  indicates  that the
moduli problems defined by both equations are equivalent\footnote{In fact, to solve the
 complete moduli problem \cite{dijk}, one must add the
 following equations for the curvatures: $ d_A \Phi_i = 0$, $ [\Phi_i, h^I] = 0$, $[\Phi_i, \Phi_j] = 0 $.}.

Expanding the action $S$, and integrating out $T$ and $H^I$,  reproduces 
the $\mathcal{N}= 4$ action in its twisted form
 \cite{Vafa,MQ}\footnote{As a matter of fact, if the manifold on
 which the theory is defined is not hyperK\"{a}hler, the $J^I$ can
 not be considered as constant. In this case the scalar fields $h^I$
 acquires a ``mass'' term linear in the selfdual part of the Riemann tensor.}
\begin{multline}
S 
\approx \int_M \trace \Biggl( - \frac{1}{2} F \star F
 + 
 \frac{1}{4}
d_A h_I \star d_A h^I + 2 d_A \Phi^i 
\star d_A \Phi_i \Biggr . - 2 \chi^\alpha_I J^I \star d_A
\Psi_\alpha + 2 \Psi^\alpha \star d_A
\eta_\alpha \\*+ 2 \star \eta^\alpha [ 
h_I, \chi^I_\alpha] + J_I \star \Psi^\alpha [ h^I , \Psi_\alpha] +
\star \varepsilon_{IJK} \chi^{\alpha\, I} [ h^J, \chi^K_\alpha] \\* - 2
\star \sigma^{i\, \alpha\beta} \chi_{\alpha\, I} [ \Phi_i,
\chi^I_\beta] - 2 \star \sigma^{i\, \alpha\beta} \eta_\alpha [\Phi_i,
\eta_\beta] - 2 \sigma^{i\, \alpha\beta} \Psi_\alpha \star [ \Phi_i,
\Psi_\beta] \\*- \frac{1}{8} \star [ h_I, h_J] [ h^I, h^J] - 2 \star [ \Phi^i ,
h_I] [\Phi_i, h^I] - 4 \star [ \Phi^i, \Phi^j][\Phi_i, \Phi_j] \Biggr) 
\end{multline}

In fact, it was not necessary 
 to ask $SL(2,\mathds{R})$-invariance from the beginning. Rather, looking  for a  $\delta$, $s$ and $\bar s$ invariant action, with internal symmetry $SU(2)\times U(1) $,  (U(1) is the ghost number symmetry), determines a    unique action, Eq.~({\ref{action}), with additional    $SL(2,\mathds{R})$
   and $\bar \delta$ invariances. This shows that the $\mathcal{N}=
 4$ supersymmetric action is determined by the invariance under the action of only 6 generators $s,\bar s, \delta$, with a much smaller internal symmetry than the $SL(2,\mathds{H})$ R-symmetry, namely  $SU(2)\times U(1) $.  After untwisting,  this determines 
 the  relevant  closed off-shell sector of $\mathcal{N}=4$
 supersymmetry.
 
 
\subsection{Horizontality condition in four dimensions}
The algebra is not
contained in a single horizontality condition, as in the
seven-dimensional case.  
In fact, one has splitted  conditions   for the Yang--Mills field, and   for the scalar fields $h^I$. (This     gives the possibility of building matter multiplets, by relaxing the condition that $h^I$ is in the adjoint representation of the gauge group). They are :
\begin{gather}
(d + s + \bar s + \delta + \bar \delta) \scal{ A + c + \bar c +
 |\kappa| \gamma + |\kappa| \bar \gamma } + \scal{
 A+ c + \bar c + |\kappa| \gamma + |\kappa| \bar \gamma }^2 
 \hspace{20mm}\CR \hspace{10mm}
= F + \Psi + \bar\Psi + g(\kappa) \scal{\eta + \bar \eta}+ g(J_I
\kappa) \scal{\chi^I + \bar \chi^I} + ( 1 + |\kappa|^2 ) \scal{ \Phi +
 L + \bar \Phi } 
\CR[2mm]
(d_A + s_{\scriptscriptstyle (c)} + \bar s_{\bar c} + \delta_\gamma + \bar \delta_{\bar
 \gamma}) h^I = d_A h^I + \bar\chi^I - \chi^I + i_{J^I \kappa}
\scal{\bar\Psi - \Psi} 
\\*[4mm]
(d + s + \delta - i_\kappa) \scal{ A + c + |\kappa| \gamma} + \scal{
 A+ c + |\kappa| \gamma}^2 
= F + \Psi + g(\kappa) \eta + g(J_I \kappa) \chi^I + \Phi + |\kappa|^2
\bar \Phi 
\CR[2mm]
 (d_A + s_{\scriptscriptstyle (c)} + \delta_\gamma - i_\kappa) h^I = d_A h^I + \bar\chi^I +
i_{J^I \kappa} \bar\Psi 
\end{gather}
These equations and their Bianchi identities fix  the action of $s$, $\bar s$, $\delta$ and $\bar \delta$, by expansion in ghost number, up to
the introduction of auxiliary fields that are needed for solving  the indeterminacies of
the form ``$s$ antighost + $\bar s$ ghost'' . 
 These indeterminacies
introduce   auxiliary fields in the equivariant part of the algebra,
$T$ and $H^I$,  as well as  in  the Faddeev--Popov sector, with several auxiliary fields that  induce  a breaking of the $SL(2,\mathds{R})$ invariance. The latter    does not affect the equivariant, that is, gauge invariant, sector.
To be more precise about the number of auxiliary fields, all the actions given by a symmetrized product of operators
on the four Faddeev--Popov ghosts are determined by the closure
relations of the algebra. There is one indeterminacy for
each antisymmetrized product of operators. To
close the algebra in the Faddeev--Popov sector,   $11=6+4+1$  Lagrange multiplier fields must be introduced, with the standard technique.   We do not give here the complete algebra for these fields, which we postpone   for  a further paper, devoted to a  new  demonstration of the
finiteness of the $\mathcal{N}=4,D=4$ theory.

We can gauge-fix the action in a $s$ and $\bar s$ invariant way {\it
 and/or} in a $s$ and $\delta$ invariant way. In the former case,  one uses
an $s \bar s$-exact term which gauge-fixes the connection $A$.. This $s \bar s$-exact term eliminates all  
  fields of the Faddeev--Popov sector, but $c, \, \bar c$, and $b
  \equiv s \bar c$.  In the former case,  one uses 
an $s \delta$-exact term. In this case, $\bar c$ is replaced   by $ \gamma$.  

\section{Different twists of $\mathcal{N}= 4$ superYang--Mills}
As noted in \cite{MQ,t3} there are three non equivalent twists of
$\mathcal{N}=4$ superYang--Mills, corresponding to  the different possible
choices of an $SU(2)$ in the R-symmetry group $SL(2,
\mathds{H})$. When one defines the symmetry by horizontality
conditions, these 3 different possibilities correspond to different
representations of the matter fields. 
These matter fields are respectively organized in an $SL(2,
\mathds{R})$-Majorana--Weyl spinor, a vector field and a
quaternion. The latter is the one  studied in the previous
section \footnote{The field $L$ is the real part of the quaternion
  and the fields $h^I$ the imaginary one.},  and, as a matter
of fact, the most studied in the literature
\cite{dijk,Vafa,wang,and}. It is the only case    that can be
understood   as a dimensional reduction of the
eight-dimensional   topological theory. We will shortly see that the two  other cases have    scalar and vector symmetries that are not big enough for a determination of the action.  

\subsection{Spinor representation}
The first twist gives an $\mathcal{N}_T=1$ theory. It is obtained by
breaking $Spin(4) \otimes SL(2, \mathds{H})$ into 
$Spin(4)\otimes SU(2) \otimes SL(2,\mathds{R}) \otimes U(1)$, and then   taking  the diagonal of the chiral $SU(2)$ of $Spin(4)$ with the
$SU(2)$ of the previous decomposition of   $SL(2, \mathds{H})$. 
The bosonic matter field $h^\alpha$ is then a chiral $SL(2,
\mathds{R})$-Majorana--Weyl spinor. The ghost $\lambda^\alpha_+$ and
antighost $\lambda^\alpha_-$ of the matter field are respectively
chiral and antichiral $SL(2, \mathds{R})$-Majorana--Weyl spinors. 
The horizontality condition reads:
\be
(d_A + s_{\scriptscriptstyle (c)} + \delta_{\scriptscriptstyle (c)} - i_\kappa ) h^\alpha = d_A h^\alpha +
\lambda^\alpha_+ + \ba \kappa \lambda^\alpha_-
\ee
Introducing   the auxiliary antichiral $SL(2,
\mathds{R})$-Majorana--Weyl spinors $D^\alpha$, one gets : \begin{align}
s_{\scriptscriptstyle (c)} h^\alpha &= \lambda^\alpha_+ \hspace{10mm} &\delta_{\scriptscriptstyle (c)} h^\alpha
&= \ba \kappa \lambda^\alpha_- \CR
s_{\scriptscriptstyle (c)} \lambda^\alpha_+ &= [\Phi, h^\alpha] &\delta_{\scriptscriptstyle (c)}
\lambda^\alpha_+ &= \ba \kappa D^\alpha + \Lc_\kappa h^\alpha \CR
s_{\scriptscriptstyle (c)} \lambda^\alpha_- &= D^\alpha &\delta_{\scriptscriptstyle (c)} \lambda^\alpha_- &=
\ba \kappa [ \bar \Phi, h^\alpha ] \CR
s_{\scriptscriptstyle (c)} D^\alpha &= [\Phi, \lambda^\alpha_-] &\delta_{\scriptscriptstyle (c)} D^\alpha &=
\ba \kappa [ \eta, h^\alpha ] + \ba \kappa [ \bar \Phi,
\lambda^\alpha_+] + \Lc_\kappa \lambda^\alpha_-
\end{align} 
With this definition of the twist,  there is no other scalar or vector
charge, which leaves us  with a $\mathcal{N}_T= 1$ theory. The action
is not completely determined by these two symmetries. 

\subsection{Vector representation}
One breaks  $SL(2, \mathds{H})$ into $Spin(4) \otimes SO(1,1)$
and then takes the diagonal of $Spin(4) \otimes Spin(4)$ \cite{marcus}. 
 The matter horizontality condition involves a vector field $V^\mu\equiv h^\mu $,
  its vector ghost $ \bar \Psi$ and antighosts scalar $\bar \eta$
  and  selfdual  2-form $\bar \chi$,  
\be
(d_A + s_{\scriptscriptstyle (c)} + \delta_{\scriptscriptstyle (c)} - i_\kappa) V = d_A V + \bar \Psi +
g(\kappa) \bar \eta + i_\kappa \bar \chi
\ee
This give the following transformations of the fields
\begin{align}
s_{\scriptscriptstyle (c)} V &= \bar \Psi \hspace{10mm} &\delta_{\scriptscriptstyle (c)} V &= g(\kappa) \bar\eta +
i_\kappa \bar \chi \CR
s_{\scriptscriptstyle (c)} \bar \Psi &= [\Phi, V] &\delta_{\scriptscriptstyle (c)} \bar\Psi &= g(\kappa) h +
i_\kappa \bar H + \Lc_\kappa V \CR
s_{\scriptscriptstyle (c)} \bar\eta &= h &\delta_{\scriptscriptstyle (c)} \bar\eta &= -[\bar\Phi, i_\kappa V] \CR
s_{\scriptscriptstyle (c)} h &= [\Phi, \bar \eta] &\delta_{\scriptscriptstyle (c)} h &= [\eta , i_\kappa V] -
[\bar\Phi, i_\kappa \bar \Psi] + \Lc_\kappa \bar\eta \CR
s_{\scriptscriptstyle (c)} \bar\chi &= \bar H &\delta_{\scriptscriptstyle (c)} \bar\chi &= - 2 [\bar\Phi,
\scal{g(\kappa) V}^+] \CR
s_{\scriptscriptstyle (c)} \bar H &= [\Phi, \bar\chi] &\delta_{\scriptscriptstyle (c)} \bar H &= 2 [ \eta,
\scal{g(\kappa) V}^+] - 2[\bar\Phi, \scal{g(\kappa) \bar\Psi}^+] +
\Lc_\kappa \bar\chi
\end{align}
This corresponds to a $\mathcal{N}_T = 2$ theory. However, the mirror
operators $\bar s_{\scriptscriptstyle (c)}$ and $\bar
\delta_{\scriptscriptstyle (c)}$ have the same ghost 
number as the primary ones. The internal  symmetry in this case is  
  $\mathds{Z}_2$ instead of $SL(2,\mathds{R})$  \cite{MQ}.  As a
  matter of fact the four 
symmetries are not enough to fix the action and do not give an algebra
which can be closed off-shell without the introduction of an infinite
set of fields.

\subsection*{Acknowledgments}

This work was partially supported under the contract ANR(CNRS-USAR) \\ \texttt{no.05-BLAN-0079-01}.



\begin{thebibliography}{99}


\bibitem{BBT}
 L.~Baulieu, G.~Bossard and A.~Tanzini,
 ``Topological Vector Symmetry of BRSTQFT and Construction of Maximal
 Supersymmetry,''
 \texttt{[arXiv:hep-th/0504224]}.
 
 
 \bibitem{renor}
 V.~E.~R.~Lemes, M.~S.~Sarandy, S.~P.~Sorella, A.~Tanzini and O.~S.~Ventura,
 ``The action of $N = 4$ super Yang--Mills from a chiral primary operator,''
 JHEP {\bf 0101}, 016 (2001)
 \texttt{[arXiv:hep-th/0011001]};
 N.~Maggiore and A.~Tanzini,
 ``Protected operators in $\mathcal{N} = 2,\, 4$ supersymmetric theories,''
 Nucl.\ Phys.\ B {\bf 613}, 34 (2001)
 \texttt{[arXiv:hep-th/0105005]}.

\bibitem{moore}
 R.~Dijkgraaf and G.~W.~Moore,
 ``Balanced topological field theories,''
 Commun.\ Math.\ Phys.\ {\bf 185}, 411 (1997)
 \texttt{[arXiv:hep-th/9608169]}.

\bibitem{dijk}
 R.~Dijkgraaf, J.~S.~Park and B.~J.~Schroers,
 ``$\mathcal{N} = 4$ supersymmetric Yang--Mills theory on a K\"{a}hler
 surface,'' 
 \texttt{[arXiv:hep-th/9801066]}.

\bibitem{glouk}
 D.~Mulsch and B.~Geyer,
 ``$G_2$ invariant 7D Euclidean super Yang--Mills theory as a
 higher-dimensional analogue of the 3D super-BF theory,''
 Int.\ J.\ Geom.\ Meth.\ Mod.\ Phys.\ {\bf 1} (2004) 185
 \texttt{[arXiv:hep-th/0310237]}.

\bibitem{Vafa}
 C.~Vafa and E.~Witten,
 ``A Strong coupling test of $S$ duality,''
 Nucl.\ Phys.\ B {\bf 431} (1994) 3
 \texttt{[arXiv:hep-th/9408074]}.


\bibitem{MQ}
 J.~M.~F.~Labastida and C.~Lozano,
 ``Mathai--Quillen formulation of twisted $\mathcal{N} = 4$
 supersymmetric gauge theories in four dimensions,'' 
 Nucl.\ Phys.\ B {\bf 502} (1997) 741
 \texttt{[arXiv:hep-th/9702106]}.

\bibitem{wang}
 P.~Wang,
 ``A Suggestion for Modification of Vafa--Witten Theory,''
 Phys.\ Lett.\ B {\bf 378} (1996) 147
 \texttt{[arXiv:hep-th/9512021]}.

\bibitem{matter}
 N.~Dorey, V.~V.~Khoze and M.~P.~Mattis,
 ``Multi-instanton calculus in $\mathcal{N} = 2$ supersymmetric gauge
 theory. II: Coupling to matter,''
 Phys.\ Rev.\ D {\bf 54} (1996) 7832
 \texttt{[arXiv:hep-th/9607202]}.

\bibitem{t3}
 J.~P.~Yamron,
 ``Topological Actions From Twisted Supersymmetric Theories,''
 Phys.\ Lett.\ B {\bf 213} (1988) 325.

\bibitem{marcus}
 N.~Marcus,
 ``The Other topological twisting of $\mathcal{N}=4$ Yang--Mills,''
 Nucl.\ Phys.\ B {\bf 452} (1995) 331
 \texttt{[arXiv:hep-th/9506002]};
 M.~Blau and G.~Thompson,
 ``Aspects of $\mathcal{N}_T \geqslant 2$ topological gauge theories
 and D-branes,'' 
 Nucl.\ Phys.\ B {\bf 492} (1997) 545
 \texttt{[arXiv:hep-th/9612143]}.
 
\bibitem{and}
 J.~M.~F.~Labastida and C.~Lozano,
 ``Duality in twisted $\mathcal{N} = 4$ supersymmetric gauge theories in four
 dimensions,''
 Nucl.\ Phys.\ B {\bf 537} (1999) 203
 \texttt{[arXiv:hep-th/9806032]};
 M.~Jinzenji and T.~Sasaki,
 ``$\mathcal{N} = 4$ supersymmetric Yang--Mills theory on
 orbifold-$T^4 \bigr / \mathds{Z}_2 $,'' 
 Mod.\ Phys.\ Lett.\ A {\bf 16} (2001) 411
 \texttt{[arXiv:hep-th/0012242]}, 
 JHEP {\bf 0112} (2001) 002
 \texttt{[arXiv:hep-th/0109159]}.
  
\end{thebibliography}
\end{document}